\documentclass[iop]{emulateapj}

\usepackage{amsmath}
\usepackage{natbib}
\usepackage{graphicx}

\def\gtwid{\mathrel{\raise.3ex\hbox{$>$\kern-.75em\lower1ex\hbox{$\sim$}}}}
\def\ltwid{\mathrel{\raise.3ex\hbox{$<$\kern-.75em\lower1ex\hbox{$\sim$}}}}
\def\\{\hfil\break}

\def\sun{\hbox{$\odot$}}
\def\earth{\hbox{$\oplus$}}
\def\lesssim{\mathrel{\hbox{\rlap{\hbox{\lower4pt\hbox{$\sim$}}}\hbox{$<$}}}}
\def\gtrsim{\mathrel{\hbox{\rlap{\hbox{\lower4pt\hbox{$\sim$}}}\hbox{$>$}}}}

\def\arcdeg{\hbox{$^\circ$}}

\newcommand{\unit}[1]{\ifmmode \:\mbox{\rm #1}\else \mbox{#1}\fi}
\newcommand{\sbr}[1]{_{\rm #1}}
\begin{document}

\title{Dark Matter Halos in Galaxies and Globular Cluster Populations. II:
Metallicity and Morphology}
\author{William E. Harris}
\affil{Dept.\ of Physics and Astronomy, McMaster University, Hamilton, ON L8S 4M1, Canada.}
\author{Gretchen L. Harris}
\affil{Dept.\ of Physics and Astronomy, University of Waterloo, Waterloo, ON N2L 3G1, Canada.}
\author{Michael J. Hudson\altaffilmark{1}}
\affil{Dept.\ of Physics and Astronomy, University of Waterloo, Waterloo, ON N2L 3G1, Canada.}

\email{harris@physics.mcmaster.ca}
\altaffiltext{1}{Perimeter Institute for Theoretical Physics, Waterloo, Ontario, Canada.}
\shorttitle{}
\shortauthors{Harris, Harris, and Hudson}

\begin{abstract}
An increasing body of data reveals a one-to-one linear correlation between galaxy halo mass
and the total mass in its globular cluster (GC) population, $M_{GCS} \sim M_h^{1.03 \pm 0.03}$ 
valid over 5 orders of magnitude.  In this paper
we explore the nature of this correlation for galaxies of different morphological types, and for the subpopulations
of metal-poor (blue) and metal-rich (red) GCs.  For the subpopulations of different metallicity we find
$M_{GCS}(blue) \sim M_h^{0.96 \pm 0.03}$ and $M_{GCS}(red) \sim M_h^{1.21 \pm 0.03}$ with similar scatter.
The numerical values of these exponents can be derived from the detailed behavior of the 
red and blue GC fractions with galaxy mass and provide a self-consistent set of relations.  
In addition, all morphological types (E, S0, S/Irr)
follow the same relation, but with a second-order trend for spiral galaxies to have a slightly
higher fraction of metal-rich GCs for a given mass.
These results suggest that the amount of gas available for GC formation at high redshift was
in nearly direct proportion to the dark-matter halo potential, in strong contrast to the markedly 
nonlinear behavior of total stellar mass versus halo mass.

Of the few available theoretical treatments that directly discuss the formation of
GCs in a hierarchical merging framework, we find that the model of \citet{kravtsov_gnedin05} best matches these
observations.  They find that the blue, metal-poor GCs formed in small halos at $z > 3$ and did so in nearly direct
proportion to halo mass.  
Similar models addressing the formation rate of the red, metal-richer GCs in the same detail and continuing to
lower redshift are still needed for a comprehensive picture.
\end{abstract}


\maketitle

\label{firstpage}

\section{Introduction}

A remarkably simple empirical correlation has emerged between two global properties of galaxies that belong to
early phases of galaxy formation:  these two quantities are $M_h$ (the
total \emph{halo mass} of a galaxy including all visible and dark matter
in its potential well); and $M_{GCS}$ (the total mass in its \emph{globular cluster system}). Recent studies
linking globular clusters (GCs) with halo mass 
\citep[e.g.][among others; see Section 2 below for a more complete literature survey]{blakeslee_etal1997, spitler_forbes09, 
georgiev_etal10, hudson_etal2014} have shown that the simple ratio $\eta \equiv M_{GCS}/M_h$ is virtually constant
across 5 orders of magnitude in galaxy luminosity or mass.  This behavior is 
quite unlike the strongly nonlinear shape of the
more commonly known correlation of total stellar mass $M_{\star}$ with $M_h$
\citep[e.g.][]{leauthaud_etal2012, hudson_etal2015, velander_etal2014, kravtsov_etal2014}.  
On strict observational grounds, the new result indicates that there is at least one
stellar subpopulation that formed in direct proportion to galaxy halo mass.

In our previous discussion \citep[][hereafter HHH14 or Paper I]{hudson_etal2014}, we constructed a new calibration of
$\eta = M_{GCS}/M_h$ by using a recent catalog of $419$ GC systems (GCSs) in all types of galaxies
\citep[][hereafter HHA13]{harris_etal13}.  The GCS data were combined
with a single, homogeneous calibration of $M_{\star}$ versus $M_h$ founded solely   
on the extensive CFHLens weak lensing program \citep{velander_etal2014, hudson_etal2015}. 
Measurements of $M_{\star}$ in turn 
were determined from $K-$band (or $V-$band in cases where $K$ is unavailable)
luminosities of the galaxies along with appropriate color-dependent mass-to-light ratios.
The resulting correlation, built from 307 galaxies covering all galaxy 
types and all environments, confirms that $M_{GCS}$ varies almost exactly one-to-one with
$M_h$.  The residual scatter is $\pm0.35$ dex, and arguments 
taking into account the different weights of the individual $M_{GCS}$ values and an
estimated $\gtrsim 0.15$ dex uncertainty in $M_h$ suggest that the intrinsic scatter of the 
correlation may be no larger than $\pm 0.2$ dex.

In HHH14 we treated all galaxies alike, and also treated their GCs as a monolithic 
population regardless of age or metallicity.  In the present paper, we pursue the
empirical correlation further by (a) subdividing the GC samples in each galaxy into their
astrophysically interesting ``red'' and ``blue'' metallicity groups, 
and (b) considering subgroups of different morphological type (ellipticals, S0's, and S/Irr types).
We then look for any differences among these subsets that may provide
additional clues for GC formation.

In Section 2, we provide a more thorough overview of the observational and theoretical 
background for this topic, while in Section 3 we discuss some important features of the
database.  In Section 4 we describe the $M_{GCS}-M_h$ correlation broken 
down by galaxy type and by GC metallicity groups.  In Section 5 we discuss possible interpretations,
and in Section 6 we summarize the results.

Readers interested primarily in the newer results and the analysis might find it useful to skim
Section 2 on the background of the observations and relevant theory, skip Section 3 on the
treatment of the database, and then go to Sections 4 and 5 for the results.

\section{Background on the GCS-$M_h$ Correlation}

The idea that the GC population might scale directly with the total mass in its
host galaxy has arisen several times over the past two decades.
However, the previous literature on this topic is sparse enough that a fairly complete assessment of it
can be made.  To date, it has been an observationally driven subject.  Thus we 
look first at the previous papers drawing on empirical data, and next
at the (even fewer) papers directly addressing theoretical motivations for the correlation.

\subsection{The Observational Material:  Growth of the Database}

The first quantitative appearance of the correlation
is in \citet{blakeslee1997}, \citet{blakeslee_etal1997}, and \citet{blakeslee1999}, who measured the GCSs in 
a sample of two dozen giant galaxies in Abell clusters.   Several of these are
BCGs (Brightest Cluster Galaxies), dominant objects within their environments.  
Blakeslee and colleagues suggested in essence that the total
GC population $N_{GC}$ scaled linearly with the total (baryonic + dark)
mass of their host galaxies as measured by halo X-ray temperature or velocity dispersion
\citep[see Fig.~15 and accompanying discussion of][]{blakeslee_etal1997}.
To quote from their paper, ``These observations might be explained by a model in which the GCs ... 
formed early on and in proportion to the available mass''.
However, their sample was restricted to a small number of giant galaxies in unusually
rich environments and at that stage the generality of the statement was not clear.

\citet{mclaughlin1999} pursued a roughly similar line with a larger
sample of 97 galaxies covering the size range from dwarfs to giants, focussing
instead on the ratio of $N_{GC}$ to the total estimated \emph{baryonic mass}
$M_{bary}$ in a galaxy.  In dwarf galaxies a large fraction of the initial gas mass has been
lost due to feedback, and in supergiants, a large fraction of the \emph{initial} baryonic
mass is now in X-ray-luminous halo gas. If we add the
extra assumption that the initial value of $M_{bary}$ is proportional to $M_{h}$, i.e. 
$N_{GC} \sim M_{bary}(init) \sim M_h$, the argument connecting GCs to halo mass
becomes similar in nature.  Lastly, \citet{kavelaars1999} used 11 large galaxies to show that
$N_{GC}$ scaled well with the \emph{dynamical} mass of the galaxy within $\sim 20$ kpc,
as calculated from the velocity dispersion of the GCS itself.

A decade later, a new thread of this empirical argument
was started by \citet{spitler_etal2008} using a selected sample of 25 large and nearby
galaxies. For this sample, as in \citet{blakeslee1997} they showed that the ratio of
GC \emph{number} to halo mass $(N_{GC}/M_h)$ was nearly constant, but now extended to galaxies over a wider
range of environments and masses.   
\citet{spitler_forbes09} discussed similar material for a larger sample of $\sim 100$ galaxies, and
changed the thrust of the argument from GC numbers to GCS \emph{mass} versus $M_h$.
This difference is not trivial, because mean GC mass $\langle M_{GC} \rangle$ 
increases systematically with galaxy luminosity; see \citet[][]{villegas2010} and HHA13.
The calibration of $M_{h}$ in Spitler \& Forbes used a combination of two studies for
weak-lensing masses of isolated galaxies, and (for lower-luminosity galaxies) conditional
luminosity function analysis.  

\citet{peng_etal08} used the GCS data from their HST survey of $\sim 100$ Virgo cluster galaxies
to discuss both the luminosity fraction (denoted $S_L$) and mass fraction (denoted $S_M$) 
of GCSs per unit \emph{stellar} mass $M_{\star}$. 
From these, they showed that the assumption $N_{GC} \sim M_h$ can lead
to the well known very non-linear trend of $S_M$ versus galaxy stellar mass.
\citet{georgiev_etal10} assembled $N_{GC}$ totals for a still wider
range of galaxies drawn from their own work on dwarf galaxies and
from several other sources in the literature, along with galactic stellar and gas masses.
They then defined the specific GCS mass as $S_M = M_{GCS}/M_{bary}$ where now
$M_{bary} = M_{\star} + M_{gas}$, analogous to the ratio used by \citet{mclaughlin1999}. The distinction between
this definition and the one by \citet{peng_etal08} (who used $S_M = M_{GCS}/M_{\star}$)
is significant for very gas-rich
dwarfs and for giant ellipticals with massive amounts of X-ray halo gas.
To translate $S_M$ into $\eta = M_{GCS}/M_h$ and to
calibrate $M_h$, they used a combination of dynamical mass
measurements of dwarfs and galaxy group dynamics, though in a partially model-dependent way.
For their sample of galaxies they find roughly Gaussian-shaped distribution 
functions for $\eta$ with small mean differences by galaxy type.

HHA13 compiled all the $N_{GCS}$ data available in the literature (419
galaxies at time of publication, containing many more galaxies at high luminosities
than in previous discussions).  This material was used to generate 
newly constructed values of $M_{GCS}$, explicitly accounting for the dependence of
mean GC mass on galaxy luminosity.  They combined these with 
calculated dynamical masses for the galaxies, $M_{dyn} = const~\sigma_e^2 R_e$.
Since $M_{dyn}$ is computed at the effective radius $R_e$ within which stellar  
matter dominates, it is a reasonable proxy for $M_{bary}$ or $M_{\star}$.  Like the previous authors, they find
that the mass ratio $(M_{GCS} / M_{dyn})$ correlates nonlinearly with galaxy luminosity in
a way that is consistent with a constant $\eta$ across the full range of galaxies.

In HHH14 we took two additional steps:  (a) the near-infrared luminosity of each galaxy
$M_K$ was used along with mass-to-light ratios from \citet{bell_etal03}
to represent $M_{\star}$, and (b) $M_{\star}$ was translated to $M_h$ with a
single, homogeneous calibration based entirely on weak lensing data \citep{hudson_etal2015}.

More recently, \citet{forte_etal2014} have taken 67 galaxies from the Virgo HST/ACS survey,
divided them into 9 luminosity bins, and constructed a `fiducial' galaxy for each bin that
represents the average properties of the galaxies in the bin.  They use these fiducials to
derive the mean trend of $N_{GC}$ per unit dark-matter halo mass for both the
blue and red GC subsystems, finding nonlinear trends in opposite senses.
Their calibration of $M_h$ follows the interpolation functions of \citet{shankar_etal2006}
based on occupation index and numerical simulations.

A question of special interest is the presence of intracluster GCs within rich clusters of 
galaxies, which seem to belong to the overall potential well of the galaxy cluster as a whole and
center on the BCG.  These types of GCs have been found in Virgo \citep{lee_etal2010} and
Coma \citep{peng_etal11} and appear to be predominantly metal-poor, consistent with an 
origin from stripped satellite galaxies.  
\citet{durrell_etal2014} present new results from the NGVS (Next Generation Virgo Cluster Survey)
for the GC populations across the entire Virgo region.  They derive an
estimate of $\eta$ applicable to an entire galaxy cluster, where a large fraction of
the total number of GCs includes intracluster light.
In this case, the dark-matter mass in Virgo is
derived from X-ray halo gas temperatures and satellite dynamics.  

In Table 1, we summarize the different estimates of $\eta$ published in these observationally
based papers and the number of galaxies used in each.  The mutual agreement 
for $\eta$ is within a factor of two.  The source of the 
differences from one study to the next is partly the steadily increasing size of 
the GCS databases used by each study in turn, but more importantly it is due to the differences in the
detailed assumptions about key parameters in the problem:  
the conversion of $N_{GC}$ into $M_{GCS}$, the mean
GC mass, the method for defining $M_{\star}$, and the conversion of $M_{\star}$ into $M_h$.
These various assumptions are summarized in the last three columns of Table 1.

\begin{table*}
\begin{center}
	\caption{\sc Measurements of $\eta = M_{GCS}/M_h$}
	\label{tab:eta}
	\begin{tabular}{llllll}
		\tableline\tableline\\
		\multicolumn{1}{l}{Source} &
		\multicolumn{1}{l}{n} &
		\multicolumn{1}{l}{$\eta~(10^{-5})$} &
		\multicolumn{1}{l}{$\langle M_{GC} \rangle$} &
		\multicolumn{1}{l}{$M_{\star}$} &
		\multicolumn{1}{l}{$M_h$} \cr
		\\[2mm] \tableline\\
		Blakeslee etal.~1997 & 21 & 17: & $2.4 \times 10^5 M_{\odot}$ & $L_V$ & X-ray gas \\
		Spitler et al.~ 2008 & 25 & 3.2 & $2.4 \times 10^5 M_{\odot}$ & $L_K, L_V$ & weak lensing \\
		Spitler \& Forbes 2009 & 100: & 7.1 & $4 \times 10^5 M_{\odot}$ & $L_K, L_V$ & lensing + conditional LF\\
		Georgiev et al.~2010 & 200: & $6 \pm 1$ & $1.69 \times 10^5 M_{\odot}$ & $L_V +$ gas & model, dynamics \\
		Harris et al.~2013 & 255 & 6.0 & function of $M_V^T$ & $\sigma_e^2 R_e$ & Lensing + abundance matching \\
		Hudson et al.~ 2014 & 307 & $3.9 \pm 0.9$ & function of $M_V^T$ & $L_K, L_V$ & weak lensing \\
		Durrell et al.~2014 & -- & $2.9 \pm 0.5$ & $2.4 \times 10^5 M_{\odot}$ & $L_V$ & X-ray, satellite dynamics \\
		This study & 175 & $3.4 \pm 0.4$ & function of $M_V^T$ & $L_K, L_V$ & weak lensing \\
		\\[2mm] \tableline
	\end{tabular}
\end{center}
{\sc Columns:} (1) Literature source for the $\eta-$value. \\
(2) Number of galaxies used. \\
(3) Derived value of mass ratio $\eta$. \\
(4) Assumed mean GC mass needed to convert $N_{GC}$ into $M_{GCS}$.  For three studies (HHA13, HHH14, and the present work),
both the mean GC mass $\langle M_{GC} \rangle$ and the dispersion of the Gaussian GCLF are functions of galaxy luminosity
$M_V^T$.  Details of the calculation are in HHA13. \\
(5) Method for deriving stellar mass $M_{\star} \simeq M_{bary}$. \\
(6) Method for converting stellar mass to DM halo mass. 
\vspace{0.4cm}
\end{table*}

\subsection{Theoretical Interpretations}

The history of theoretical models relevant to the $M_{GCS} - M_h$ correlation
is even shorter.  \citet{diemand_etal2005} and \citet{moore_etal06} used high-resolution
cosmological N-body simulations to conclude that the amount of material collected in high$-\sigma$ 
peaks at the very earliest stages of star formation 
($z \sim 12$) should increase with the virial halo mass, roughly as $M_h^{1.2}$
(see Fig.~8 of Moore et al.). These peaks should be the
places where proto-GCs would form.  By hypothesis,
then the oldest \emph{metal-poor} (blue) globular cluster population should rise with 
halo mass, though perhaps nonlinearly.  Their picture, however, is closely linked to the role of reionization, which
they and others before that \citep[][]{beasley_etal02, santos2003, bekki2005}
suggested may have truncated GC formation at a slightly later stage.  In addition, 
it is not clear how the metal-rich (red) GCs should fit in to this picture.

Nearly simultaneously, \citet[][hereafter KG05]{kravtsov_gnedin05} used a
high-resolution hydrodynamic simulation to
explore the formation of a typical Milky-Way-type galaxy, covering the time
period from $z = 11.8$ to 3.35 (set by limitations of computer time).  This range covers the formation
of the oldest (metal-poor, blue) GCs but unfortunately does not continue long enough to
reach the [Fe/H] $> -1$ range of the metal-rich, red GC population (see their Figure 8).
The minimum cell size of their models is $\simeq 40$ pc and thus individual GCs
are not resolved.  Instead, within each mini-halo of the merger tree,
they identify the sites of GC formation as the densest cores above a threshold density of 1 $M_{\odot}$ pc$^{-3}$
in the emerging GMCs (giant molecular clouds) 
\citep[cf.][]{harris_pudritz94, kruijssen2014}.  The input physics contains both
internal SNe heating and external cosmic UV background.  The time evolution of their model
galaxies typically shows frequent bursts accompanying mergers
and infall of gas-rich satellites, and rapid buildup of a massive central galaxy up to
the end of the model run.  The spatial distribution of the newly formed GCs
ends up more centrally concentrated than the overall halo potential, but most intriguingly
for our purposes, the total mass in the
GCs is found to track the masses of their host subhalos as $M_{GCS} \sim M_h^{1.13 \pm 0.08}$.  
This correlation is drawn from their simulated halos over a mass range 
$6 \times 10^9 M_{\odot} \lesssim M_h \lesssim 3 \times 10^{11} M_{\odot}$.

In essence, both the Moore et al.~and KG05 models should be compared more strictly with only
the blue metal-poor GC population that emerged at higher redshift.
We emphasize this comparison in Section 4 below.

In another cosmological simulation directed towards finding GCs at formation,
\citet{bekki_etal08} identified proto-GCs as the central particles within
virialized sub-halos.  The simulation covers a large enough cosmological volume that
many model galaxies can be inspected, thus building up GCS properties as functions
of galaxy size.  They find the peak rate of metal-poor GC formation to be at
$z \simeq 8 - 6$ and for metal-rich GCs at $z \sim 4$, in reasonable agreement
with Milky Way GC ages (see above).  For the ratio of $N_{GC}$ to halo mass, they find a shallow, smooth 
increase of about a factor of two going from dwarf galaxies up to giants corresponding approximately
to $N_{GC} \sim M_h^{1.3}$, though with differences depending on galaxy morphology.
A dependence this steep would not match the current data.

Recently \citet{katz_ricotti2014} have modelled the possible dependence of $\eta$ on redshift 
(time of formation) by tracing the effects of an assumed $\eta(z)$ on the present-day 
observed properties of the GCS.  Given the amount of dynamical evolution and cluster destruction
since formation, they deduce an increase of a factor of several at
higher $z$, and also a higher $\eta$ in lower-mass halos.

In both HHA13 and \citet{kruijssen2014}, the empirical result $\eta = const$ is
viewed essentially as a product of two ratios, 
$\eta \sim (M_{\star}/M_h) \cdot (N_{GC}/M_{\star})$.
Each of these ratios varies nonlinearly with $M_{\star}$, but in opposite
senses, so that when multiplied together their nonlinearities essentially cancel out.  If so, GC formation 
might be largely disconnected from the dark matter halo and the uniformity of $\eta$ is
something of an accident.  However, this view would give no explanation for the detailed
way each ratio varies with galaxy mass, or why they should be so exactly complementary.

An essential piece of the puzzle is the role of feedback external
to the proto-GCs during early star formation.  
In dwarf galaxies, internal feedback including SNe, stellar UV emission,
stellar winds, and external heating from reionization may remove much of the original
gas and truncate field-star formation.  For giant galaxies, feedback through AGNs and infall heating
will similarly prevent much of the original gas in the DM potential from forming stars.
However, the extremely dense, small proto-GCs may have been much less affected by any of these
forms of feedback and more nearly reflect
the original gas supply.  As suggested before, the essential two-step link
from halo mass to GCS mass is therefore $M_{GCS} \sim M_{bary}(init) \sim M_h$.
As also seen in the theoretical simulation studies listed above, the role of the DM halo is to set up
the conditions where dense gas collects together and the proto-GCs can form.

As noted above, several previous discussions of GC formation are intimately connected with the role of cosmic reionization.
In most of these papers, reionization is invoked to \emph{truncate} the formation
of the earliest, metal-poor GCs and thus help produce the observed bimodal metallicity
distribution of GCs \citep[e.g.][]{beasley_etal02, santos2003, bekki2005, spitler_etal2012, moran_etal2014}.
This approach would require that most of the metal-poor GCs should have formed before the
reionization peak, i.e.~in the regime $z \gtrsim 10$.
A second view is that reionization \emph{triggered} GC formation by shocking 
protocluster clouds into collapse \citep{cen01}. Still a third view is that the reionization epoch is
largely \emph{irrelevant} to GC formation, either because the protocluster clouds and their GMCs
are self-shielded \citep{muratov_gnedin10, barkana_loeb2002, shapiro_etal2004}, or
because most GCs formed after the main epoch of reionization 
\citep[e.g.][and KG05]{bekki_etal08, kruijssen2014}.
KG05 suggest particularly that the conditions needed for GC formation 
(namely, growth of very dense cores embedded within massive, cooled GMCs) could only accumulate in a large-scale
way after reionization and should have peaked around $z \sim 5$.

Reionization peaked at $z \simeq 11$ \citep{hinshaw_etal2013, ade_etal2014} and was complete near
$z \simeq 6$ \citep{fan_etal06, zahn_etal2012, jensen_etal2014}.  By comparison, 
direct measurements of GC ages that do not depend on cosmological models 
find ages for Milky Way GCs distributed rather evenly from 11 to 13 Gy, equivalent to a redshift range $z \simeq 2.5 - 7$
\citep[see, e.g.][for recent analyses]{marin-franch_etal09, hansen_etal13, leaman_etal2013, vandenberg_etal2013}.
With the standard techniques of isochrone fitting to deep color-magnitude diagrams for GCs,
internal precisions of $\pm 0.5$ Gy are now achievable,
though uncertainties of $\sim \pm 1$ Gy in the \emph{absolute} ages
remain due to small uncertainties in the different stellar model calculations, the adopted
GC distance scale, the adopted GC metallicities, and details of
isochrone fitting procedures \citep[see especially][for a thorough discussion]{vandenberg_etal2013}.

If the GC age measurements summarized above are correct, the oldest GCs formed clearly
\emph{before} the peak epoch of star formation in the universe at $1 \lesssim z \lesssim 3$
\citep{behroozi_etal13, madau_dickinson2014}, but \emph{after} reionization
had largely finished (though recognizing the \emph{caveat} in the absolute GC
age calibrations).  The reason for bringing in reionization was to produce the
universal GC bimodality, but this may also be unnecessary.  An alternate and perhaps
more attractive explanation
for bimodality emerges from the fact that the merger tree for a large galaxy contains very
few intermediate-sized halos in which intermediate-metallicity GCs would form; most mergers
are strongly unequal-mass ones where dwarfs are accreted by the central galaxy 
\citep[][]{cote_etal98, muratov_gnedin10, tonini2013}.
These arguments combined 
favor the third option that cosmic reionization and other forms of feedback had far less influence
on GC formation than on the lower-density field star population.

Though it is at early stages, the theory available so far suggests we should look 
more closely at the $M_{GCS} - M_h$ correlations for the easily identifiable
GC subpopulations (blue, red) and also host galaxy type.  We do this in section 4 below, after
some comments about the nature of the GCS database itself.

\begin{figure}[t]
	\vspace{0.0cm}
	\begin{center}
\includegraphics[width=0.5\textwidth]{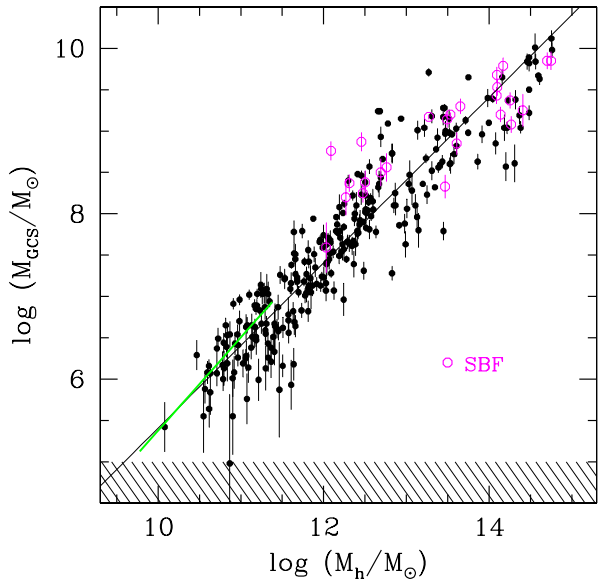}
\end{center}
	\vspace{-0.5cm}
\caption{Correlation of the total mass in globular clusters, $M_{GCS}$, versus the total
	halo mass $M_h$ of a galaxy.  Masses are in units of $M_{\odot}$.  The errorbars on each
	point show the uncertainties due to the measured total GC population in each
	galaxy. Magenta open circles are galaxies for which the GC population was measured
	through the SBF technique (see text).
	The \emph{solid line} is the least-squares solution for the ``All'' sample listed in Table 2 and
	has a slope of unity. 
	The \emph{shaded region} at bottom shows the range of
	low-mass galaxies that would on average be expected to have less than 1 GC (and thus
	would not appear in our database).
	At lower left, the \emph{green line} shows the theoretical prediction
	from \citet{kravtsov_gnedin05} over the halo mass range covered by their model.}
	\vspace{0.5cm}
\label{fig:mhalo_all}
\end{figure}

\begin{figure}[t]
	\vspace{0.0cm}
	\begin{center}
\includegraphics[width=0.5\textwidth]{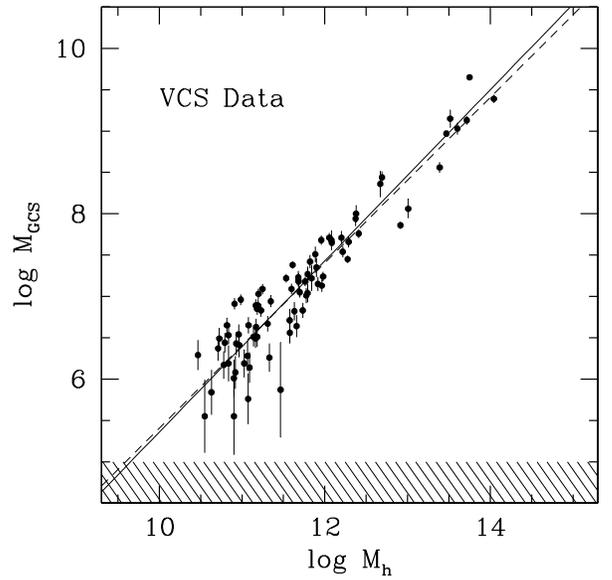}
\end{center}
	\vspace{-0.5cm}
\caption{Correlation of the total mass in globular clusters, $M_{GCS}$, versus the total
	halo mass $M_h$ of a galaxy, now for 84 galaxies in the HST-based Virgo Cluster 
	Survey \citep{peng_etal08}.  
	The \emph{dashed line} shows the best-fit solution for all galaxies, from
	Figure 1, while the \emph{solid line} shows the solution from only the Virgo
members.  The \emph{green line} at lower left is the KG05 model line as in Figure 1.}
	\vspace{0.5cm}
\label{fig:mhalo_virgo}
\end{figure}

\section{The GCS Database and Tests of Systematics}

In the present paper we continue to work with the data in 
the HHA13 catalog of globular cluster systems.  
Because the catalog is assembled from more than 100 individual GCS studies in the
literature, inhomogeneity is undoubtedly present and may be responsible for some
of the observed scatter in the basic correlation of $M_{GCS}$ versus $M_h$
\citep[e.g.][]{zaritsky_etal2014, usher_etal2013}.
However, the estimated $N_{GC}$ values 
in most cases are derived from a simple and highly standardized
method now used in almost all GCS studies.
In brief, this ``standard method'' is:
\begin{itemize}
\item{} Begin with imaging of the target galaxy that is as deep as possible, as 
wide-field as possible, and uses as many filters as possible,
\item{} Carry out photometry across the imaging field down to some objectively
defined limiting magnitude (usually the 50\% completeness level as determined
by artificial-star experiments).
\item{} Subtract the background level (field contamination).  The GCS shows up as a statistical excess
of starlike or near-starlike objects, centered on the galaxy and showing up above the uniform
background of `field' objects. The field contamination can be reduced
by a number of methods including elimination of extremely red or blue objects,
objects that have structures marking them as faint background galaxies, 
and objects that have very different scale sizes than GCs at the distance of
the target galaxy.  For HST-type resolution, the GCs may be partially resolved
and thus not starlike.  
\item{} Calculate the residual total measured GC population in the field after background subtraction. 
Correct this total for the unobserved outer-halo portion beyond the field of view of the
study, normally using the power-law or Sersic-law radial distribution of the GCs as 
measured directly from the data.
\item{} Finally, correct the total to include GCs fainter than the photometric limit
of the data. Here, we gain a large advantage from the fact that the GC luminosity
distribution (GCLF) has a standard lognormal shape in galaxies of all types and
sizes, a long-established result resting on extremely strong observational footing
\citep{villegas2010, jordan2007, harris_etal2014}.
\end{itemize}

Excellent recent examples of the application of this standardized method 
can be found in \citet{zaritsky_etal2014, brodie_etal2014, young_etal2012, harris09a, cho_etal2012}
to mention only a few.
An extreme example is the uncovering of the GCS in the nearby giant elliptical
NGC 5128 holding a system of $\simeq 1000$ GCs \citep{harris_etal2004a, harris_etal2004b,
harris_etal2012}.  This target in some sense represents the most difficult combination of all the 
selection techniques listed above:  the surrounding field has large amounts of contamination
from both foreground stars and faint background galaxies; spatial resolution is not as effective
since the GCs are marginally resolved under good ground-based seeing; and the galaxy is so nearby
that its GCS is spread across a large projected area of the sky.  Even so, reliable results for
the GC total $N_{GC}$ and GCLF have been obtained, with several studies showing mutual consistency
\citep{gharris2010}.

The most prevalent sources for any inhomogeneity in the current catalog (HHA13) are likely to be 
field size and radial coverage, which differ quite a bit from one
individual study to another \citep[see also][]{usher_etal2013,brodie_etal2014}.  
The galaxies most affected by limited field coverage are
\emph{giant, nearby} ones that have been observed with cameras with relatively small fields of
view, such as the HST WFPC2, ACS, or WFC3 detectors.  In some studies, careful work has
been done to assess the field corrections \citep[e.g.][]{peng_etal08}, even calibrated against
wide-field ground-based imaging, but the resulting uncertainties in $N_{GC}$ are inevitably larger
than for studies drawing directly on wide-field imaging.  For the limiting-magnitude correction,
most modern studies reach or exceed the GCLF peak or `turnover' point
above which half the GC population is found, and the universally consistent shape of the
GCLF makes the resulting normalization to $N_{GC}$ easy to carry out.  

The issue for the present paper in practice boils down to two questions:  (a) does inhomogeneity
\emph{systematically} affect either the slope or zeropoint of the correlation; and (b) does it generate significantly
increased \emph{scatter}?  An empirical test of both questions can be made with the
study in the literature that probably has the strongest claim to internal homogeneity,
the HST-based Virgo Cluster Survey \citep[VCS;][]{peng_etal08}.  Here all target galaxies are at
virtually the same distance and were observed with the same camera (ACS/WFC), the same exposure
times, and the same filters. Even here, however, field size is still an issue because a smaller
fraction of the total GCS falls within the field of view for larger galaxies.  

The basic correlation of $M_{GCS}$ versus $M_h$ including all galaxies with measurements
of both $N_{GC}$ and $M_h$ is shown in Figure 
\ref{fig:mhalo_all}, and for comparison 
Figure \ref{fig:mhalo_virgo} shows the correlation for only the 84 Virgo members with both halo
masses from HHH14 and $N_{GC}$ numbers adopted from \citet{peng_etal08}.\footnote{We note here that
	Virgo contains two supergiant ellipticals of similar visual luminosity, M49 and M87.  
	M87 is more likely to be the Virgo BCG because it is sitting near the center of the large-scale
	Virgo potential well and is surrounded by an extended halo of hot X-ray gas, intracluster
field stars, and intracluster GCs \citep{durrell_etal2014}.}
Encouragingly, the best-fit solution from the Virgo data alone 
is not significantly different from the solution from all catalog data.
However, the scatter is certainly lower (formally 0.30 dex for Virgo, versus 0.35 dex for all data).

The entire catalog has many more galaxies than does the Virgo sample
in the mass range $M_h > 10^{12} M_{\odot}$, including
all but one of the BCGs, and it is in this high-mass range that the scatter is perhaps most problematic.
Though the inhomogeneity of the data for the larger galaxies may well be the cause of the
increased scatter, it is hard to state this definitively since
the Virgo cluster contains only a few such galaxies, so it may also be possible
that some of the scatter is intrinsic.  A single, more homogeneous observational survey
targeted specifically to a large sample of high-mass galaxies would provide a key test of this
point \citep[see][]{zaritsky_etal2014} as well as being valuable in its own right.

For each distinct subset of the data we carry out a separate best-fit solution
from a weighted fit of all data points.  Specifically, we assume
that data pairs $(x_i, y_i)$ with measurement uncertainties $(\sigma_{x,i}, \sigma_{y,i})$ 
are related by 
\begin{equation}
	y \, = \, \alpha \, + \, \beta(x - \langle x \rangle)
\end{equation}
and then minimize the $\chi^2$ estimator \citep[][]{novak_etal2006, gharris_etal2014}
\begin{equation}\label{equation-chi2}
	\chi^2 \, = \, \sum{ {(y_i - \alpha - \beta (x_i - \langle x \rangle))^2} \over {(\sigma_{y,i}^2 + \epsilon_y^2)
    + \beta^2 (\sigma_{x,i}^2 + \epsilon_x^2) } } \, .
\end{equation}
In this case, $x \equiv {\rm log} M_h$ and $y \equiv {\rm log} M_{GCS}$. 
In Eq.~\ref{equation-chi2}, $\epsilon_x$ and $\epsilon_y$ denote any intrinsic or ``cosmic"
scatter present in the data and additional to the direct measurement uncertainties. 
Without extra knowledge of some kind about the data, we would not be able to gauge
both $\epsilon_x$ and $\epsilon_y$ independently \citep[see][]{gharris_etal2014}.

Table 2 lists the results for all the datasets discussed in this paper.
Following HHH14 we set $\sigma_x = 0.15$ dex
for all points, whereas $\sigma_y$ is determined by the uncertainties in $N_{GC}$ as given in the
catalog.  For the initial solutions listed in Table 2 we set $\epsilon_x = \epsilon_y = 0$, lacking
extra information that would specify otherwise.  The last column of the Table gives the total rms
scatter of the residuals in $y$.
For all the solutions listed, the fits were performed iteratively removing any objects with
$y-$residuals larger than 1.0 dex.  The $n-$values listed in column 3 give the number of galaxies
remaining for the final fit.

\begin{table*}
\begin{center}
	\caption{\sc Observational Correlation of $M_{GCS}$ with $M_h$}
	\label{tab:correlation}
	\begin{tabular}{lllllll}
		\tableline\tableline\\
		\multicolumn{1}{l}{Galaxy Type} &
		\multicolumn{1}{l}{GC Subset} &
		\multicolumn{1}{l}{$n$} &
		\multicolumn{1}{l}{$\alpha (\pm)$} &
		\multicolumn{1}{l}{$\beta (\pm)$} &
		\multicolumn{1}{l}{$\langle x \rangle$} &
		\multicolumn{1}{l}{$\sigma$ (rms)} \cr
		\\[2mm] \tableline\\
		All Types & All & 293 & 7.706 (0.020) & 1.000 (0.020) & 12.3 & 0.349 \\
		          & Blue & 167 & 7.405 (0.025) & 0.960 (0.027) & 12.2 & 0.323 \\
		          & Red  & 145 & 7.157 (0.030) & 1.209 (0.030) & 12.2 & 0.355 \\
		          & No SBF & 269 & 7.598 (0.021) & 1.004 (0.019) & 12.2 & 0.347 \\
			  & $f_{red}$ known (``Best'') & 166 & 7.564 (0.024) & 1.026 (0.025) & 12.1 & 0.306 \\
		          & Virgo only & 82 & 7.118 (0.033) & 1.035 (0.043) & 11.7 & 0.302 \\
		\\ \tableline \\
		E & All & 171 & 8.041 (0.027) & 0.960 (0.023) & 12.6 & 0.348 \\
		  & Blue & 97 & 7.549 (0.035) & 0.945 (0.036) & 12.2 & 0.340 \\
		  & Red  & 86 & 7.321 (0.041) & 1.186 (0.041) & 12.3 & 0.376 \\
		\\ \tableline \\
		S0 & All & 77 & 7.627 (0.037) & 1.150 (0.049) & 12.2 & 0.326 \\
		   & Blue & 47 & 7.253 (0.037) & 0.980 (0.056) & 12.0 & 0.254 \\
		   & Red  & 38 & 7.063 (0.045) & 1.304 (0.037) & 12.2 & 0.280 \\
		\\ \tableline \\
		S/Irr & All & 46 & 7.086 (0.052) & 0.985 (0.078) & 11.8 & 0.350 \\
		      & Blue & 24 & 7.171 (0.068) & 0.825 (0.085) & 12.0 & 0.334 \\
		      & Red & 21 & 6.913 (0.070) & 1.107 (0.114) & 12.0 & 0.322 \\
		\\[2mm] \tableline
	\end{tabular}
\end{center}
\vspace{0.4cm}
\end{table*}

The HHA13 catalog contains about 40 galaxies for which $N_{GC}$ was measured with the SBF 
(surface brightness fluctuation) technique \citep{blakeslee_etal1997, marin-franch2002}
where the GCS is detected through the imprint of the unresolved GCs on the power spectrum
of the integrated light, rather than the standard method outlined above.
These objects are plotted separately in Fig.~1, and in Table 2 we list the solutions both
including and excluding them.  These two solutions are the same within the uncertainties
of the coefficients ($\alpha, \beta$), but the SBF technique does not fall into the `standard method'
category and in what follows we exclude any data from this method.

\begin{figure}[t]
	\vspace{0.0cm}
	\begin{center}
\includegraphics[width=0.5\textwidth]{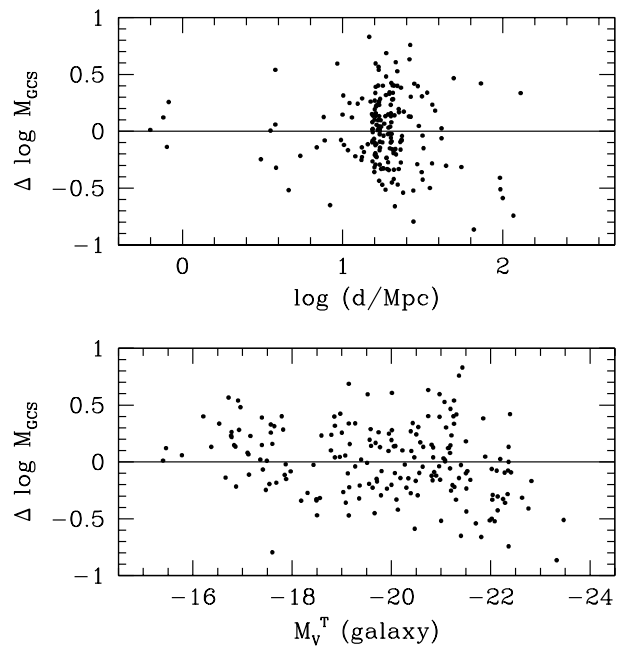}
\end{center}
	\vspace{-0.5cm}
	\caption{\emph{Upper panel:} Residuals (measured minus predicted) from the best-fit correlation
		of $M_{GCS}$ versus $M_h$, plotted versus galaxy distance. The sample shown includes only galaxies with well
		determined GC populations (uncertainties $< 0.15$ dex) and excludes any measured with the SBF
		technique (see text).  
	\emph{Lower panel:} Residuals plotted versus absolute visual magnitude of the galaxy.}
	\vspace{0.5cm}
\label{fig:resid_d}
\end{figure}

To investigate two other possible systematic effects,
in Figure \ref{fig:resid_d} we show the residuals from the $M_{GCS} - M_h$ 
correlation plotted versus galaxy distance $d$, and versus galaxy luminosity $M_V^T$.
The data shown are for objects with the most well determined GCS totals
($< 0.15$ dex in $N_{GC}$) and also exclude objects measured only with the SBF
technique mentioned above.
The halo masses are determined from the integrated near-IR luminosities
$K_{tot}$, transformed into $M_{\star}$ and then into $M_h$ through the
calibration curve described in HHH14.  Thus \emph{if} the luminosities of large, nearby
galaxies are underestimated because of aperture-size limitations, these galaxies
would show positive residuals since their predicted halo
masses would be too small.  No obvious effect of this type shows up in Figure 3a.  
Similarly, in Figure 3b there is no strong trend with galaxy luminosity. 
As discussed in \citet{spitler_forbes09} and HHH14, for many of the BCGs particularly in large
Abell clusters, $M_{GCS}$ is
likely to be underestimated because their GC populations are so spatially extended that
they encounter field-of-view limits in most observational studies.  Individual corrections
for field size and intracluster populations must be made and are difficult to predict.
To date, these specific
corrections have been made only for the BCGs in Virgo \citep[M87; see][]{durrell_etal2014}
and Coma \citep[NGC 4874; see][]{peng_etal11}.  Contrarily, $M_h$ is
not as affected because the prescriptions for defining halo masses in central galaxies such as BCGs 
\citep{hudson_etal2015} include their extended halos.  The net result would be to leave
negative residuals, which is what we see for many of the
highest-mass galaxies in Fig.~\ref{fig:mhalo_all}.  Most of these are missing in the
more highly selected sample of Fig.~\ref{fig:resid_d}.
In the only two cases
(Virgo, Coma) where it has been possible to perform a wider-scale census of the GCs in the
entire cluster, this negative bias is also reduced.

In summary, we have no evidence that use of the entire GCS catalog is preventing us from
finding the necessary global trends.  With the BCG-related problem in mind, it appears worthwhile to use the entire
GCS catalog of data to investigate other astrophysically interesting correlations, with
GC metallicity and galaxy type.  These are discussed in the next sections.

\section{The Mass Ratios versus Metallicity and Morphology}

\subsection{Cluster Metallicity:  Blue vs.~Red}

In most galaxies, the GC metallicity or color distribution has a bimodal shape with 
the blue (metal-poor) and red (metal-rich) modes
often partially overlapping.  The population fractions are $f_{red} = N(red)/N(tot)$ and $f_{blue} = 1 - f_{red}$.
With the assumption that
the mass distribution functions for both blue and red GCs are the same, then we also have
$f_{red} = M_{GCS}(red)/M_{GCS}$ 
\citep[see][]{villegas2010, ashman_etal1995, tamura_etal2006, larsen_etal2001}.
The blue mode is centered near $\langle$Fe/H$\rangle$ $\simeq -1.5$ and the red mode 
near $\langle$Fe/H$\rangle$ $\simeq -0.5$, though the peaks of both modes show 
second-order increases with galaxy luminosity \citep{brodie_strader06, peng_etal06, forte_etal2009}.
The dividing line between the blue and red modes is conventionally defined
as the central minimum in the color or metallicity histogram, which usually lies near [Fe/H] $\simeq -1.0$
as it does in the Milky Way.  

We have inspected the literature sources for all galaxies in the HHA13
catalog and wherever possible extracted the fractions $f_{red}$. These values will be listed in the next
catalog release.  In many cases the individual studies
already list $f_{red}$, usually derived from a double-Gaussian fit to the color or metallicity
distribution.  In a few papers, we have reconstructed them from the available data.
For almost half the list, however, the red/blue ratios could not be determined 
because the photometric studies were carried out in single filters,
or the sample size was too small, or the color indices were not precise enough to 
identify bimodality clearly.  Any GCSs measured through only the SBF method are also excluded.
The net result is a list of $f_{red}$ values for 219 catalog galaxies, 
of which 175 have calculated halo masses $M_h$ and 14 are BCGs.  We label this the ``best'' subsample
of catalog data.\footnote{The ``best'' sample is defined as the galaxies for which the source photometry was 
good enough to determine the red and blue GC fractions; though these are not necessarily the ones with the
smallest nominal errorbars on $N_{GC}$.}

\begin{figure}[t]
	\vspace{0.0cm}
	\begin{center}
\includegraphics[width=0.5\textwidth]{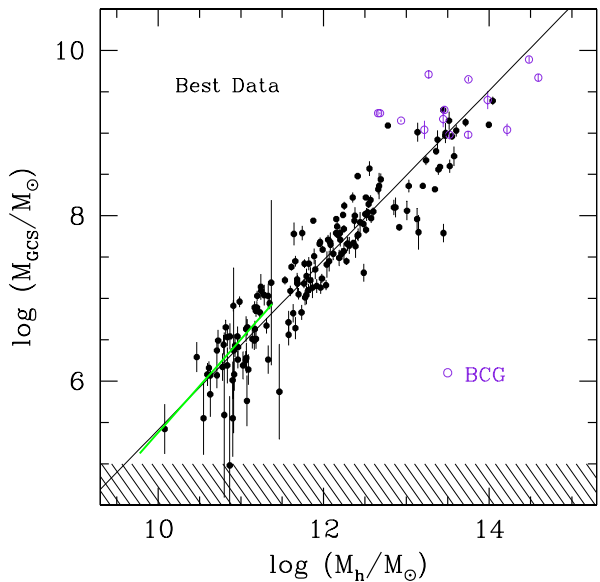}
\end{center}
	\vspace{-0.5cm}
\caption{Correlation of $M_{GCS}$ with$M_h$ for the ``best'' subsample of 175 galaxies
	as defined in the text.  BCGs are shown as the magenta open symbols at upper right,
	and the best-fit line has a slope of $1.03 \pm 0.03$.  The green
line and shaded region at bottom are as defined in Fig.~\ref{fig:mhalo_all}.} 
	\vspace{0.5cm}
\label{fig:best}
\end{figure}

\begin{figure}[t]
	\vspace{-2.0cm}
	\begin{center}
\includegraphics[width=0.5\textwidth]{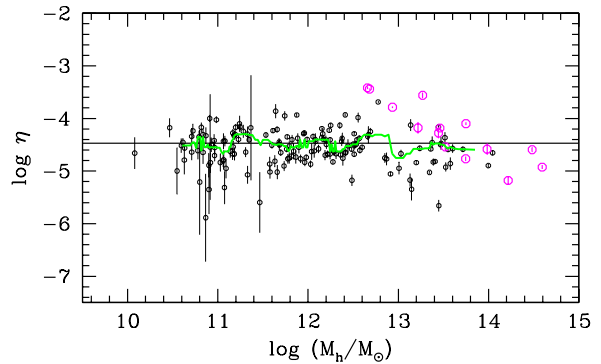}
\end{center}
	\vspace{-0.5cm}
	\caption{Mass ratio of the GCS to the halo mass $\eta$ versus halo mass, for the
	175 selected galaxies with best data (see text).  BCGs are shown as magenta open circles.
	The green line marks a running median value of $\eta$ calculated in bins of 12 galaxies each.}
	\vspace{0.5cm}
\label{fig:eta}
\end{figure}

The $M_{GCS} - M_h$ correlation for this selected dataset is shown in Figure \ref{fig:best},
for which we find a best-fit solution $M_{GCS} \sim M_h^{1.03 \pm 0.03}$.  The residual scatter is
0.3 dex, similar to the quality of fit for the Virgo subsample (Fig.~\ref{fig:mhalo_virgo}) but with the advantage
that the higher-luminosity range of galaxies is more thoroughly covered.
Interestingly, many of the BCG and other giant galaxies that had negative residuals
in Fig.~\ref{fig:mhalo_all} have dropped out of Fig.~\ref{fig:best}. 
Some of these large, distant galaxies had lower-quality data that prevented blue and red fractions 
from being determined reliably,
and some were measured with the SBF technique that automatically excludes them from the
red/blue plot.  
The distribution of $\eta$ for these 175 galaxies is shown in Figure \ref{fig:eta}.  The weighted-mean
value of the mass ratio is $\langle {\rm log} \eta \rangle = -4.47 \pm 0.07$ (listed in
the last line of Table 1).  

In Fig.~\ref{fig:eta}, we also show a running median value of $\eta$ in
bins of 12 galaxies each, where the successive bins are stepped through the list sorted by $M_h$. 
The median does not deviate far from $\eta \simeq const$, and we conclude for
the present that the simple linear form $M_{GCS} \sim M_h$ cannot be ruled out by 
the data.  The range of most concern is probably the high-mass end $M_h \gtrsim 10^{13} M_{\odot}$, 
where most of the points fall below the global average.  However, too much
significance should not be put on these biggest, highest-mass points for at least two reasons:
first, a slight change in the adopted
power-law slope of the conversion between $M_{\star}$ and $M_h$ at the high end \citep{hudson_etal2015} 
would remove this mean offset entirely. Second, as mentioned above and discussed by HHH14, 
the estimated GC populations
for many of these tend to be underestimated because of the huge spatial extent of their 
GC systems, which smoothly merge into whatever intracluster GC population may be present \cite[e.g.][]{durrell_etal2014}.

\begin{figure}[t]
	\vspace{0.0cm}
	\begin{center}
\includegraphics[width=0.5\textwidth]{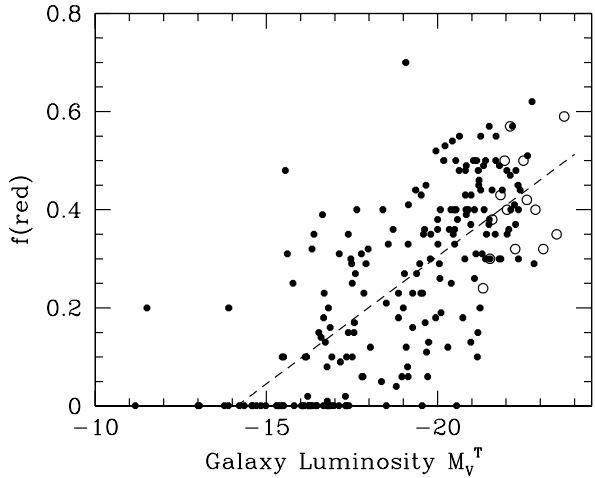}
\end{center}
	\vspace{-0.5cm}
	\caption{Fraction of the GC population that is metal-rich (red), plotted versus
	host galaxy luminosity $M_V^T$.  BCGs are plotted as open circles.  The mean
correlation given in the text is shown with the dashed line.}
	\vspace{0.5cm}
\label{fig:fred}
\end{figure}

The correlation of $f_{red}$ with galaxy luminosity $M_V^T$ is shown in Figure \ref{fig:fred}.
More luminous galaxies tend to have higher fractions of metal-rich GCs
\citep{peng_etal06}, but with considerable scatter.  Many dwarfs have only
blue GCs present ($f_{red} = 0$, at bottom of figure).  The mean trend is very roughly    
given by $f_{red} \simeq 0.25 - 0.052(M_V^T + 19)$, with residual scatter
equal to $\pm0.13$ rms.\footnote{If we use log $M_{\star}$ on the x-axis instead of $M_V^T$, the distribution is
very much the same and with the same scatter, but with fewer datapoints because the $K_{tot}$ IR
luminosities are not available for many galaxies.}

\begin{figure}[t]
	\vspace{0.0cm}
	\begin{center}
\includegraphics[width=0.5\textwidth]{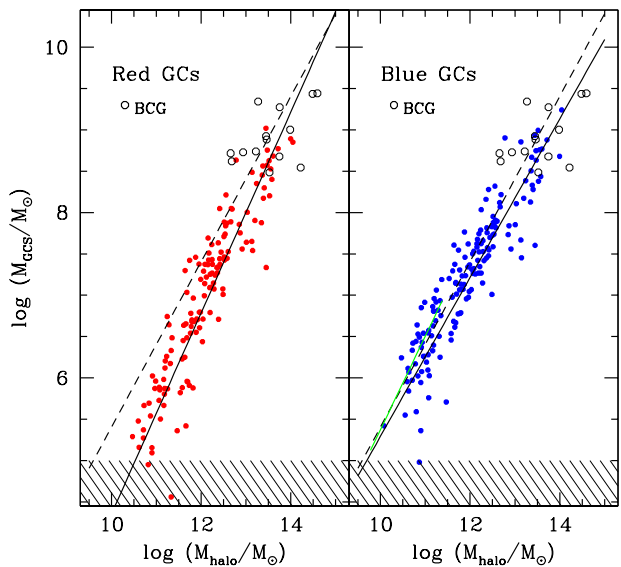}
\end{center}
	\vspace{-0.5cm}
	\caption{\emph{Left panel:} Total GC mass in the metal-rich GCs versus halo mass.
	\emph{Right panel:} Total GC mass in the metal-poor GCs.  The KG05 model prediction
	is shown as the green line at lower left.  In both panels the best-fit
solution from Table 2 is shown as the solid line, while the solution for all GCs combined 
is the dashed line.}
	\vspace{0.5cm}
\label{fig:redblue}
\end{figure}

\begin{figure}[t]
	\vspace{0.0cm}
	\begin{center}
\includegraphics[width=0.5\textwidth]{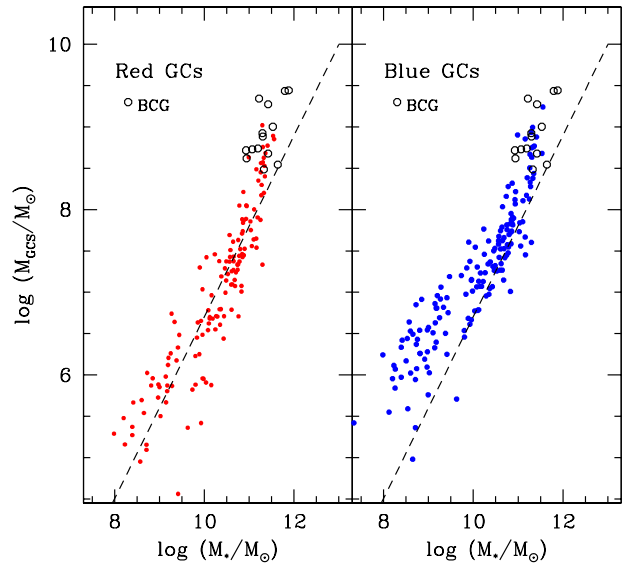}
\end{center}
	\vspace{-0.5cm}
\caption{Correlation of the total mass in globular clusters, $M_{GCS}$, versus the 
	stellar mass $M_{\star}$ of a galaxy. Symbol types are as in Fig.~\ref{fig:redblue}.
	The \emph{dashed line} in each panel shows the expected relation for galaxies
with a GC specific frequency $S_N = 1$ (see text).}
	\vspace{0.5cm}
\label{fig:redbluelight}
\end{figure}

The large scatter is partly intrinsic, but inspection of the source literature also
indicates that part of it arises from differing and incomplete field coverage between studies, as discussed above.
Because the bluer GCs are spatially more extended around the host galaxy than the red ones,
any study covering only the inner regions of the halo will overestimate $f_{red}$ unless the radial
distributions are explicitly corrected for 
\citep[these corrections are done, e.g., in][]{peng_etal08, harris09b, durrell_etal2014}.  Nevertheless, it is worth noting
that in 86\% of all galaxies in the sample, the majority of the GCs 
even from the uncorrected totals are the blue, more metal-poor ones, and even in many of
the supergiants.  Galaxies where the metal-richer GCs dominate are quite rare, and these few are not necessarily
the most massive galaxies.  For the central giant galaxies particularly, many of their
outer-halo GCs should be from captured dwarf satellites that mostly contained metal-poor GCs.
The intracluster GC population considered by itself is largely metal-poor \citep{lee_etal2010, durrell_etal2014}.

Figure \ref{fig:redblue} shows the results for $M_{GCS}(red) = f_{red} M_{GCS}$ and 
$M_{GCS}(blue) = f_{blue} M_{GCS}$ plotted as before versus $M_h$.  
For comparison, Figure \ref{fig:redbluelight} shows the same data plotted versus $M_{\star}$.
In Fig.~\ref{fig:redblue}, note that the KG05 model line is displayed only for the blue-GC subgroup, since the
metallicity range it covers does not apply to the red subgroup.  
Linear least-squares solutions
for each are shown as the solid lines and listed in Table 2.  For these solutions, we adopt an
extra dispersion $\epsilon_y = 0.1$ given the uncertainties mentioned above in defining $f_{red}$. 
For $M_{GCS}(red)$ we deliberately exclude 
any dwarfs with $f_{red} = 0$, to concentrate on the trends for galaxies
massive enough to have generated metal-rich GCs.  
Earlier versions of these trends using smaller sample were shown by \citet{spitler_etal2008} versus both $M_h$ and $M_{bary}$, 
and by \citet{forte_etal2014} for $N_{GC}$ versus $M_h$. 
Forte et al.~used the Virgo galaxies binned by luminosity; we prefer to show all galaxies
individually to gain a clearer idea of the scatter and relative quality of the correlation.

As expected given the dominance of blue GCs in most galaxies
(Fig.~\ref{fig:fred}), the blue slope $M_{GCS}(blue) \sim M_h^{0.96}$ is only slightly shallower
than the $M_{GCS} \sim M_h^{1.03}$ slope for all GCs. The steeper slope 
$M_{GCS}(red) \sim M_h^{1.21}$ tracks the more rapid rise of the 
metal-richer subpopulation during the enrichment history of the galaxies concerned, 
with a hint of nonlinearity as well in the mid-range of galaxy masses.
Notably, the \emph{scatters} for both subpopulations are
similar to each other and to the scatter we found for the combined population (rms $\pm 0.32-0.36$ dex).

The galaxy mass at which $f_{red}$ or
$M_{GCS}(red)$ rather suddenly plunges to zero is near $M_h \sim 10^{11} M_{\odot}$ or
equivalently $M_{\star} \sim 10^9 M_{\odot}$.  In physical terms,
\emph{this threshold represents the minimum halo mass
capable of reliably generating, and holding, GCs with metallicities [Fe/H] $\gtrsim -1$.}

In Fig.~\ref{fig:redbluelight} the \emph{dashed line} shows the expected relation
for galaxies with specific frequency $S_N = 1$ \citep{harris_vandenbergh81}, where 
$S_N \equiv N_{GC} \cdot 10^{-0.4 (M_V^T + 15)}$ and $N_{GC}$ has been converted to
$M_{GCS}$ following the relations in HHA13.  Neither relation can be adequately matched
by a single power-law slope over the entire range of galaxy size
(this is the classic ``specific frequency problem'').
The red GCs are, however, closer to varying linearly with $M_{\star}$ than are the blue
GCs, as expected if their formation epoch was closer to the universal peak of star formation
around $z \sim 2$ that produced most of the galaxies' stellar mass.
There is a restricted
intermediate-luminosity range near $10 \lesssim \rm{log} M_{\star} \lesssim 10.7$ 
(blue GCs) or $9 \lesssim \rm{log} M_{\star} \lesssim 10.7$ (red GCs) where it is
possible to claim that GCS mass is directly proportional to stellar mass.  This is precisely
the range over which $S_N$ is constant for all types
of galaxies (see Fig.~10 of HHA13). As noted earlier, $M_{\star}$ is not a true proxy for
total baryonic mass in many galaxies \citep[see][]{mclaughlin1999, georgiev_etal10} and it
would be desirable to assemble a more complete contemporary listing of $M_{bary}$ for the galaxies in
the catalog.

In short, we find that
\emph{galaxy halo mass is a much simpler predictor of GC population than galaxy
stellar mass, regardless of GC metallicity}.  

The Kravtsov/Gnedin (KG05) model prediction is plotted in Figs.~\ref{fig:mhalo_all} and
\ref{fig:redblue}b at lower left.  
The agreement in both slope and zeropoint with the real data from the dwarf galaxies is striking.
Most importantly, the model slope of $1.13 \pm 0.08$ is encouragingly close to the observational
result of $1.03 \pm 0.03$ for the ``Best'' sample defined above.  
The zeropoint agreement should, however,
be regarded as more nearly fortuitous:  the model data are
the estimated protocluster (dense core) masses at formation, whereas the observations give the \emph{present-day}
values of $M_{GCS}$, which are likely to be $\sim 2-3$ times lower than their initial values after
early mass loss and a Hubble time of dynamical evolution.  In addition, the KG05 halo masses are
evaluated at $z > 3$, while the observationally based numbers we use here for $M_h$ are the halo masses at $z \simeq 0$. 

\begin{figure}[t]
	\vspace{0.0cm}
	\begin{center}
\includegraphics[width=0.5\textwidth]{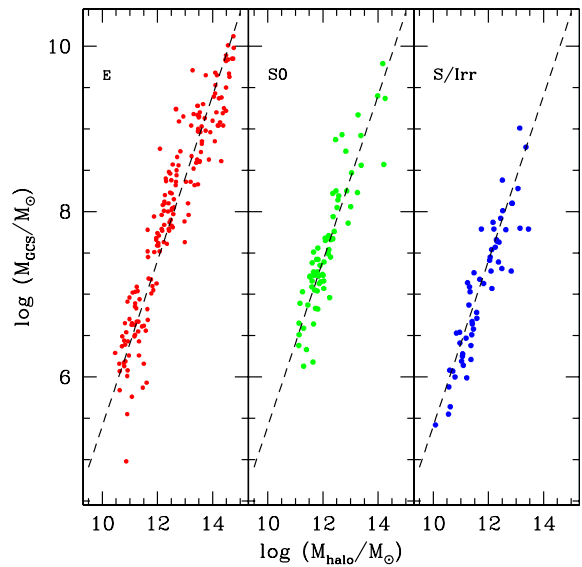}
\end{center}
	\vspace{-0.5cm}
\caption{Correlation of the total mass in globular clusters versus galaxy halo mass,
	for different galaxy morphological types.  There are 179 data points for ellipticals
	(left), 78 for S0's (center), and 46 for spirals and irregulars (right).
	In each panel the dashed line shows the mean correlation for all types combined,
from Figure 1. }
	\vspace{0.5cm}
\label{fig:type}
\end{figure}

\subsection{Galaxy Type}

An entirely new correlation that can be derived from our database is the trend
of $M_{GCS}$ with $M_h$, plotted separately for three basic morphological types:
ellipticals ($-6 \le T \le -3$), S0's ($-2 \le T \le 0$), and spirals and irregulars
together ($T > 0$).  The data are displayed in Figure \ref{fig:type} and best-fit solutions
listed in Table 2 as before.  
Here also the BCGs are not marked separately, though all of these belong to
the E group.  The scatter is smallest for the S0's, likely because many of
those are drawn from the internally more homogeneous Virgo and Fornax surveys.

\begin{figure}[t]
	\vspace{0.0cm}
	\begin{center}
\includegraphics[width=0.5\textwidth]{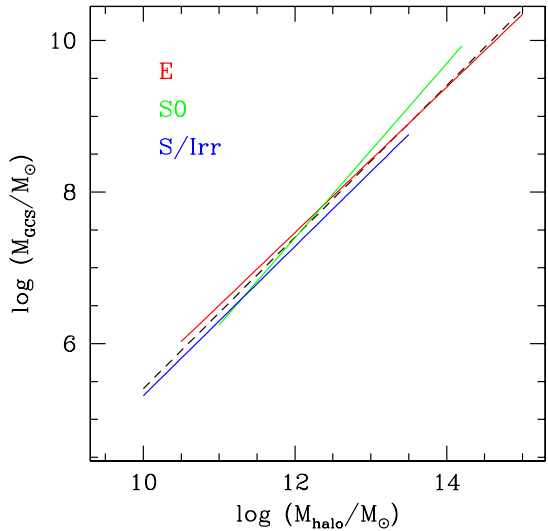}
\end{center}
	\vspace{-0.5cm}
	\caption{Mean relations from Table 2 for three different galaxy morphological
		subgroups (E in red, S0 in green, S/Irr in blue).  The dashed line is the solution for all
	galaxies combined, as in the previous Figure.}
	\vspace{0.5cm}
\label{fig:type1}
\end{figure}

In Figure \ref{fig:type1} we show the best-fit mean lines for all three galaxy types on the same
graph, along with the solution for all galaxies combined for intercomparison.  Each line is
plotted over the range of masses occupied by the available data for that subgroup.  
Though the slope for the S0's is highest by a noticeable
margin ($\beta = 1.15 \pm 0.05$ versus $1.03 \pm 0.03$ for all galaxies), the difference
is not strong and it is difficult to know what, if any, significance to ascribe to it.

A potentially more important difference -- though similarly small -- is that the mean
lines for the S/Irr and E types are nicely parallel but offset by
($0.18 \pm 0.06$) dex.  If taken at face value, this offset would mean that
the late-type galaxies were roughly 60\% as efficient as the early types in forming GCs.
Again, however, this difference is just at the margin of detectability with the current data.

An alternate explanation for the offset between the E/S0 and S/Irr mean efficiencies might be 
environmental, in the sense that early-type galaxies are found more often (though not exclusively
so) in richer groups and clusters of galaxies.  Denser environments might logically promote
higher GC and star formation efficiency at high $z$ \citep[see, e.g.][]{peng_etal08, 
kruijssen2014}.  We will explore the environmental density parameter more quantitatively
in a subsequent paper.

\section{Discussion}

There is no guarantee that a new set of data will help decipher the history of the
systems we are studying.  But the remarkably tight and structurally simple link
between GCS mass and galaxy halo mass suggests to us that it can
add to our understanding especially of 
GC formation and more generally their environments at time of formation.

\subsection{Galaxy Morphology}

The message from subdivision by galaxy type (Fig.~\ref{fig:type}) seems straightforward.
All major types follow much the same relation as defined by 
the entire sample, though with a possible second-order offset between ellipticals
and spirals.  The standard Hubble sequence was not well
established till $z \lesssim 2$ \citep[e.g.][]{mortlock_etal2013, fiacconi_etal2015}, clearly after the main
era of GC formation.  The simplest conclusion 
is that the majority of the GCs formed at an early
enough time that their conditions of formation were not strongly sensitive to the
later emergent galaxy type as we now see it. 
This evidence reinforces other striking similarities
in GC populations that have long been known.  Across all types and sizes of galaxies, 
GCs have (a) the same lognormal-type luminosity distribution  
with second-order trends
\citep[][]{jordan_etal2007, harris_etal2014}, (b) similar bimodal metallicity distributions
\citep{brodie_strader06}, and (c) the individual GCs have King-type structures and characteristic
scale radii that also change little with environment or even galactocentric
distance \citep[e.g.][]{georgiev_etal10, jordan_etal05, jordan_etal2007}.
The uniformity of the $M_{GCS}- M_h$ relation can be added to this list.
It appears to us that GC formation was controlled primarily by local hydrodynamic conditions in their parent GMCs 
and only secondarily by the much larger-scale and time-dependent structure of the protogalaxy.

\subsection{GC Metallicity}

The results from the subdivision by metallicity (Fig.~\ref{fig:redblue}) are intriguing.
Our main empirical 
finding is that \emph{GCs of both metallicity groups follow simple power-law forms
$M_{GCS} \sim M_h^{\beta}$ over the whole range of host galaxy halo mass,} quite unlike the
behavior of $M_{GCS}$ versus stellar mass for either metallicity group.

Considering the arguments from the theoretical models described above, the strong
and nearly linear correlation for the blue GCs is expected since the dwarf halos in which they formed should contain 
star-forming gas in nearly direct proportion to the halo masses, before feedback had severe effects (see above).
But the equally strong correlation for the red GCs was not necessarily expected, and current models give
less guidance for interpretation.  
The difference in slope ($\beta$(blue) = $0.96 \pm 0.03$ versus $\beta$(red) = $1.21 \pm 0.03$) is
significant, and is the direct result of the trend of $f_{red}$ versus $M_h$ (see below).

For the lower-metallicity GCs, the KG05 model gives a more quantitative way 
to understand $M_{GCS}(blue)$ versus $M_h$ over its entire range.
In a standard picture of hierarchical buildup, the majority of GCs in a final large galaxy
formed either (a) in small-to-intermediate gas-rich halos at early stages ($z \gtrsim 3$) of the merger
tree, or (b) in small satellites that were later accreted, but in which their own GCs
also formed early on.  \emph{In these small halos,}  
the KG05 model suggests that the total GC mass varies nearly one-to-one with halo mass.
Combining any number of such halos together in dry mergers without any additional
star-forming gas will then propagate the same linear proportionality 
$M_{GCS}(blue) \sim M_h$ upward to the giant-galaxy regime, automatically
replicating the correlation we now see.  

The same argument does not (yet) carry over to the metal-richer red GCs.
For these, it would not be valid simply to extrapolate the KG05 relation upward
\citep[as did][]{spitler_etal2008}, because the metal-rich GCs needed to form out of more
enriched gas in bigger halos further along the merger tree.  According to, e.g., 
\citet{muratov_gnedin10, tonini2013}, GCs with metallicities [Fe/H] $\simeq -0.5$ (the peak
of the red-GC mode) would require host halos of masses above $\sim 10^{11} M_{\odot}$,
several times larger than the upper limit of the KG05 model, in agreement with the empirical
evidence discussed in Section 4.1 above.
It would be of great interest to extend 
hydrodynamic models like KG05 to $z = 0$, in which
we should expect to see the rising proportion of metal-rich GCs as enrichment proceeds up
to near-Solar levels in the bigger halos.

Some metal-rich GCs may have formed in \emph{late, major mergers}, as can be observed
happening even at $z\simeq0$ \citep[e.g.][]{miller_etal1997, whitmore_etal1999}. But by such a late stage
the gas supply in the merger progenitors is only a small fraction of the total stellar mass already
assembled (``damp'' mergers).  
These later mergers, which depend much more strongly on the
contingent individual histories of giant galaxies, may then be responsible for adding scatter 
at the high-mass end to the overall correlations in Fig.~\ref{fig:redblue}.

\begin{figure}[t]
	\vspace{0.0cm}
	\begin{center}
\includegraphics[width=0.5\textwidth]{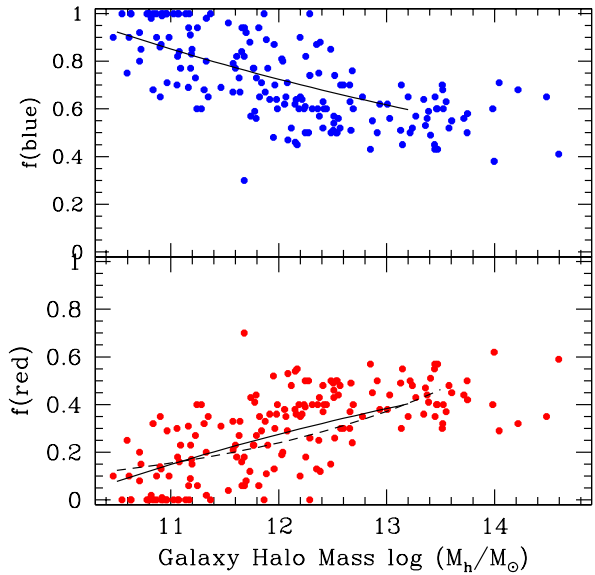}
\end{center}
	\vspace{-0.5cm}
	\caption{\emph{Upper panel:} Correlation of the blue fraction of GCs versus galaxy halo mass.  
		The solid line is the power-law fit given by Eq.~\ref{eqn:fblue} in
		the text.  \emph{Lower panel:}  Red-GC fraction versus halo mass.  The solid 
		line is the same curve as in the upper panel, converted through 
		$f_{red} = 1 - f_{blue}$, while the dashed line is $f_{red} \sim M_h^{0.19}$; see
	text for derivation.}
	\vspace{0.5cm}
\label{fig:fred_halo}
\end{figure}

To repeat, the scaling relations (Fig.~\ref{fig:redblue} and Table 2) give
$M_{GCS}(blue) \sim M_h^{0.96}$, $M_{GCS}(red) \sim M_h^{1.21}$, whereas the entire population
scales as $M_{GCS} \sim M_h^{1.03}$. 
These power-law exponents have the values they do
because of the specific behavior of $f_{red}$ or $f_{blue}$ with halo mass, which is shown in 
Figure \ref{fig:fred_halo}.  The fraction $f_{blue}$ exhibits a shallow power-law-like decline.
Suppose then that we assume $M_{GCS}(blue) \sim M_h^{\gamma}$ 
along with $M_{GCS} \sim M_h^{\beta}$ as before.  
Then since $f_{blue} \equiv M_{GCS}(blue) / M_{GCS}$, we predict 
$f_{blue} \sim M_h^{\gamma -\beta} = M_h^{-0.07}$.
This curve, given by
\begin{equation}\label{eqn:fblue}
	f_{blue} \, \simeq \, ({M_h \over {10^{10} M_{\odot}}})^{-0.07} \, ,
\end{equation}
and valid over the range $10^{10} M_{\odot} < M_h \lesssim 10^{13} M_{\odot}$,
is shown in Fig.~\ref{fig:fred_halo} (upper panel) as the solid line.
For $M_h < 10^{10} M_{\odot}$ (dwarf galaxies) we find empirically that $f_{blue} \simeq 1$ with
rare exceptions \citep{georgiev_etal10},
while for $M_h$ much above $\sim 10^{13} M_{\odot}$ the data suggest that the fractions may
saturate at around $f_{red} \simeq f_{blue} \simeq$ 0.5.  

In the lower panel of the figure, the corresponding relation for $f_{red} = 1 - f_{blue}$ is shown
as the solid line.  The dashed line in the same panel shows a simpler power-law fit that would be expected from
$f_{red} = M_{GCS}(red) / M_{GCS} \sim M_h^{1.21 - 1.03} = M_h^{0.18}$.  
Both curves provide a reasonable match within the scatter of the data. 

\begin{figure}[t]
	\vspace{0.0cm}
	\begin{center}
\includegraphics[width=0.5\textwidth]{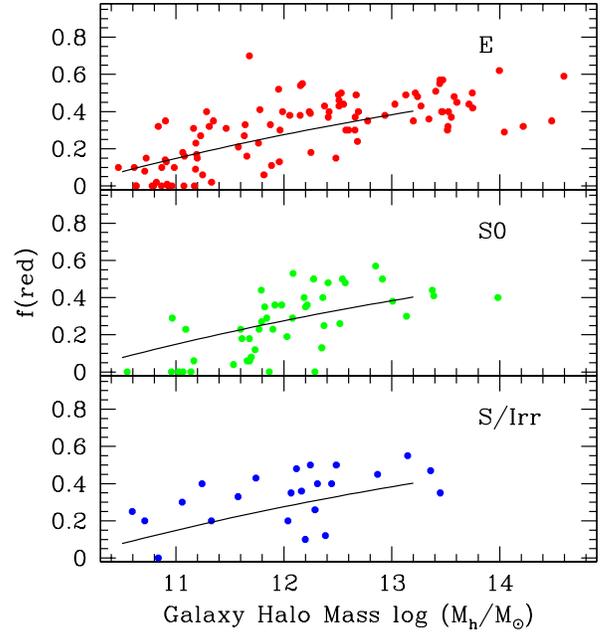}
\end{center}
	\vspace{-0.5cm}
	\caption{Correlation of the red fraction of GCs versus galaxy halo mass,
		subdivided by galaxy morphology.  
		The solid line is the power-law fit to all galaxies combined, as 
		given by Eq.~\ref{eqn:fblue} in the text and used in the previous Figure.}
	\vspace{0.5cm}
\label{fig:fred_type}
\end{figure}

Lastly, we can combine the results for the red/blue GC subcomponents with
galaxy type.  In Figure \ref{fig:fred_type} we show the growth of the red-GC fraction $f_{red}$ with galaxy
mass, subdivided into the three approximate morphological classes.  The E and S0 types behave
similarly.  But interestingly, the spiral and irregular galaxies sit slightly above the mean
line by $\Delta f_{red} \sim 0.1$, though admittedly the sample size is still
relatively small.  \citet[][]{kruijssen2014} predicts that galaxies experiencing 
fewer late mergers with small satellites should now have systematically higher $f_{red}$ values,
since they accreted relatively fewer blue, metal-poor GCs.\footnote{More specifically, the S/Irr 
	galaxies need to have accreted fewer satellites that were large enough to have globular
	clusters of their own.  The graph places no limits on the accretion of very small
satellites that may have turned into faint tidal streams but without any accompanying GCs.}  In this sense the GCSs in S/Irr
galaxies may therefore bear the imprint of a less active accretion history on average 
than the early-type galaxies experienced.

\begin{figure}[t]
	\vspace{0.0cm}
	\begin{center}
\includegraphics[width=0.5\textwidth]{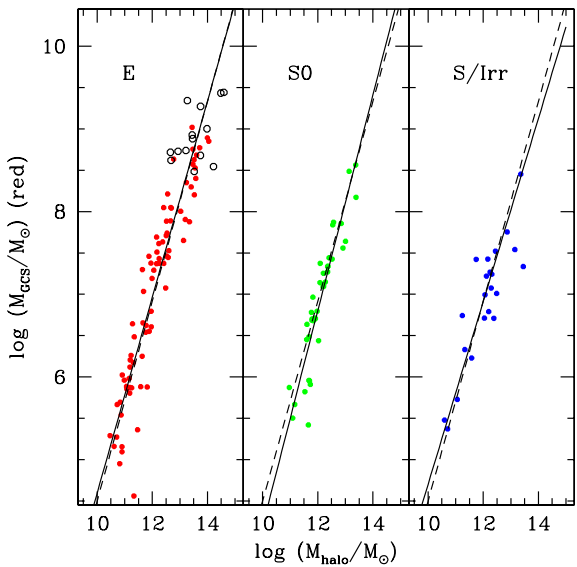}
\end{center}
	\vspace{-0.5cm}
	\caption{Correlation of the mass in red (metal-rich) GCs versus galaxy halo mass,
		subdivided by galaxy morphology.  
		The \emph{dashed line} in every panel is the fit to all galaxies combined, as 
given in Table 2.  The \emph{solid line} is the fit to each galaxy type individually.}
	\vspace{0.5cm}
\label{fig:redblue_type}
\end{figure}

In Table 2, we give the solutions further broken down into
red and blue GCs for each of the three galaxy types, and these are 
plotted for $M_{GCS}(red)$ versus $M_h$ in 
Figure \ref{fig:redblue_type}.  All three morphological groups separately agree
well with the solution for all galaxies combined.  For the S/Irr types, the higher
$f_{red}$ fraction (Fig.~\ref{fig:fred_type}) largely cancels out the lower total $M_{GCS}$ (Fig.~\ref{fig:type1}),
leaving the total mass in the red GCs similar to that in the E/S0 types.

Further discussion of these points goes beyond the scope of this paper, but we
suggest that Fig.~\ref{fig:fred_halo} and Eq.~\ref{eqn:fblue} together
represent one of the most important observational signals from 
globular cluster systems in galaxies.  
\emph{Where each galaxy lies on this graph is the visible outcome of all the GC formation events in
its individual merger tree}, in all its complexity.
A merger-tree model capable of high enough resolution to isolate GCs in formation and track
the metallicity enrichment history should
be able to reproduce this  overall trend of the blue and red GC fractions.

The intrinsic scatter present in Fig.~\ref{fig:fred_halo} around the mean relation
is partly due to measurement uncertainties, but it should also partly be due to
different individual histories for galaxies of the same
present-day mass.  Two important factors in these histories are 
(a) the numbers of dwarf satellites accreted at lower
redshift (which will boost $f_{blue}$); and (b) the number of late, major mergers
(which may boost $f_{red}$ if gas is present).  In addition, scatter may also 
arise because GC metallicity is not a strict monotonic function of formation time.   
The evidence from the Milky Way \citep{vandenberg_etal2013} is that 
metal-richer GCs are younger
by $1-2$ Gy on the average, but at any given [Fe/H] the scatter in age is at least 1 Gy. 
Conversely, at a given age the scatter in metallicity is $> 0.5$ dex. 
The formation models cited above
\citep[KG05,][]{muratov_gnedin10, griffen_etal10} typically show quick buildup of
metallicity in each subhalo, very incomplete mixing of gas, frequent merging and disruption, and 
a rapidly changing overall potential well 
till merging abates around $z \sim 2$.  Even as late as $z \sim 3$, the gas fractions within 
halos $< 10^{11} M_{\odot}$ were
well above 50\% \citep{muratov_gnedin10}, allowing GC formation in many uncorrelated starbursts
over a significant metallicity range.

\subsection{Halo and GCS Masses}

Two major criteria seem necessary to understand the basic correlation between $M_{GCS}$ and $M_h$ that
we see: \\
\noindent (1) Star formation is a function of very local conditions of gas density, temperature, and 
hydrodynamic state and is unlikely to be directly driven by the vastly larger-scale potential 
of the parent galaxy.  Thus the link connecting the two 
should be the initial gas mass collected within the parent halos, i.e.~ 
$M_{GCS} \sim M_{gas}(init) \sim M_h$. \\
\noindent (2) GCs during formation were 
relatively immune to external feedback in their larger environments including UV radiation, AGN
feedback, SNe, and stellar winds.  These feedback mechanisms may hamper or truncate field-star formation
in the large volumes of lower-density gas, but the GCs were forming in small regions where the gas density
was several orders of magnitude higher than in their host GMCs as a whole.  

If this line of reasoning is correct, 
then we need to understand better the structures of GMCs in star-forming galaxies at high redshift.
\citet{kruijssen2014} has argued that these GMCs should be more compact than present-day ones
at a given mass, with relatively more clumps of dense gas above the threshold for rapid star formation.
Thus in these, the GC formation efficiency (CFE, i.e.~the ratio of GC mass to GMC mass) may be much higher than
the present-day ratio of $\sim 1$\% used in earlier discussions \citep{harris_pudritz94}.  \citet{muratov_gnedin10} also find
that the fraction of stellar mass contained in the GCs is as high as 10\% at $z > 3$.  

On the observational side, an increasing body of measurements for star-forming galaxies at $z > 2$
is starting to show (a) that molecular gas can make up as much as half the total gas mass;
(b) that in turn, half the molecular gas can collect into GMCs capable of quickly forming stars; and
(c) the star-forming clouds themselves are denser than in the present-day universe
\citep[e.g.][]{popping_etal2014,freundlich_etal2014,debreuck_etal2014,thomson_etal2015}
\citep[see also references in][]{kruijssen2014}.

In HHH14 we proposed that the final mass ratio
$\eta$ could be thought of as a product of four separate ratios,
\begin{align}
	\eta & = \Bigl(\frac{M\sbr{bary}}{{M\sbr{h}}}\Bigr)\times
	\Bigl(\frac{M\sbr{GMC}}{M\sbr{bary}}\Bigr)\times
	\Bigl(\frac{M\sbr{PGC}}{M\sbr{GMC}} \Bigr)\times\Bigl(\frac{M\sbr{GC}}{M\sbr{PGC}}\Bigr) 
\end{align}
where $M_{bary}/M_h \simeq 0.15$ is the cosmic baryonic mass fraction; $M_{GMC}/M_{bary}$ is the fraction of \emph{initial}
gas mass that collects into GMCs big enough to build globular clusters; $M_{PGC}/M_{GMC}$ is the mass
fraction of gas in a typical GMC that ends up in proto-GCs (equivalent to
the CFE); and $M_{GC}/M_{PGC}$ is the present-day
GC mass as a fraction of its protocluster mass.

Of these four terms, only the first can be considered well known.  The fourth term can
be estimated roughly at $\simeq 0.1$ given that the star-forming efficiency in a PGC
should be $\sim 0.3-0.5$ and that the following dynamical evolution of the cluster
will reduce its mass by another factor of 3 to 5 \citep[e.g.][]{lada_lada03,katz_ricotti2014}.
The second and third ratios, however, 
are especially uncertain because we need to know them at high redshift, and because they
may well show considerable place-to-place variance.  Although far more evidence
is needed, the works cited above are starting to point to the 
view that both $(M_{GMC}/M_{bary})$ and $(M_{PGC}/M_{GMC})$ can be of order $\sim 0.1$ or higher at
the redshift range when GC formation is most active.  Conservatively, we adopt $\sim 0.05$ 
for both.

The point of writing Eq.~(4) in this form is that, even if some of the terms are speculative,
the {\sl product} of all four is well determined.  As noted in HHH14, we can then invert the question:
$\eta$ can be used to constrain the second and third terms rather than the other way around.
Our revised order-of-magnitude
estimate, setting the product of the 4 terms equal to the observed value of $\eta$, would
then be roughly
\begin{align}
	\eta & \simeq 0.15 \times 0.05  \times 0.05 \times 0.1 \simeq 4 \times 10^{-5} \, . 
\end{align}
Consistent with the observational work mentioned above, this scaling argument indicates that
several percent of all the initial gas mass
was able to collect into GMCs massive enough to build globular clusters that can survive to the present day.

Lastly, of the available theoretical models that specifically address the formation of GCs in hierarchical merging
\citep{diemand_etal2005, moore_etal06, kravtsov_gnedin05, bekki_etal08, muratov_gnedin10, griffen_etal10, tonini2013},
we find the results of KG05 to best match the data discussed here.  For the metal-poor GCs the model
successfully predicts the nearly one-to-one relation between $M_{GCS}(blue)$ and $M_h$, at least for halos
$\lesssim 3 \times 10^{11} M_{\odot}$.  If all low-metallicity GCs form in such relatively small halos, then 
later mergers to build up a larger galaxy will automatically extend the correlation upward over the entire range of galaxy masses. 
The main limitations of the model are spatial resolution (proto-GCs are simply single-pixel points of high
density); and redshift range covered ($z >3.35$).  Capturing the metal-rich GC population will require
continuing the models to $z \lesssim 2$.
Modelling capable of going beyond both those limitations is increasingly possible and should open up a new
range of tests of the theory.

\section{Summary}

We have used the database of GC system masses and the halo masses of their host galaxies to extend 
earlier discussions of the relation between these two quantities that are relics of the 
galaxy formation era.  Our main conclusions are these:
\begin{enumerate}
\item The observed linear correlation $M_{GCS} \sim M_h$ continues to become stronger with the growth of the GCS database.
	To first order it is insensitive to host galaxy 
	morphology, applying to ellipticals, S0's, spirals, and irregulars alike.  A second-order difference
	between the ellipticals and spirals may exist in the sense that the S/Irr types were slightly less
	efficient per unit mass in forming GCs.
\item The two GC metallicity modes show different behaviors with $M_h$, but both follow extremely simple
	scaling laws and similar low dispersion.  We find $M_{GCS}(blue) \sim M_h^{0.96}$, and (for galaxies
	massive enough to have generated metal-rich GCs) $M_{GCS}(red) \sim M_h^{1.21}$.  In short,
	galaxy halo mass provides an accurate predictor of both types of GC populations.  
\item The smallest dark-matter halos capable of reliably producing and holding metal-rich GCs ([Fe/H] $> -1$)
	have $M_h \simeq 10^{11} M_{\odot}$.
\item We find tentative evidence that S/Irr galaxies show a slightly higher-than-average metal-rich GC fraction $f_{red}$ 
	for a given halo mass.  This evidence may indicate that, on average, these galaxies experienced fewer
	small-satellite accretions than did E and S0 galaxies.
\item We support the view that the very large-scale halo mass and the much smaller-scale
gas-dynamical process of GC formation are linked through the
	total initial gas mass in the halo, $M_{GCS} \sim M_{gas}(init) \sim M_h$.  
\item An accompanying requirement for the observed $M_{GCS} - M_h$ correlation to make sense is then that
external feedback processes appear to have little relevance for globular cluster formation.
	In essence, the relative numbers of GCs in a galaxy may then indicate what the global star formation
	efficiency would have looked like without the influence of feedback.
\item The hierarchical model of \citet{kravtsov_gnedin05} provides a basis for understanding the observed correlation
	of the \emph{metal-poor} GCs with halo mass, namely
	$M_{GCS}(blue) \sim M_h$ over the entire range of galaxy types and masses.  Extending this
	modelling to lower redshift $z \lesssim 3$ holds out hope that the metal-richer GCs could eventually be fit into
	the same comprehensive picture.
\item The specific way in which the blue (metal-poor) and red (metal-rich) GC fractions change systematically with galaxy
	mass should provide a sensitive new constraint on hierarchical-merging models for galaxy formation
	at redshifts $z > 2$.
\end{enumerate}

\section*{Acknowledgements}

MJH and WEH acknowledge the financial support of NSERC.

\bibliographystyle{apj}
\bibliography{gc} 

\label{lastpage}

\end{document}